\documentclass[runningheads]{svmult}

\usepackage{makeidx}   % allows index generation
\usepackage[dvips,final]{graphicx}  % standard LaTeX graphics tool
\usepackage{subeqnar}  % subnumbers individual equations
\usepackage{multicol}  % used for the two-column index
\usepackage{physprbb}  % modified textarea for proceedings,
\makeindex             % used for the subject index
\def\simlt{\lower.5ex\hbox{$\; \buildrel < \over \sim \;$}}
\def\simgt{\lower.5ex\hbox{$\; \buildrel > \over \sim \;$}}
\def\degr{\hbox{$^\circ$}}
\begin{document}
\title*{Eclipse mapping of accretion discs}
\toctitle{Eclipse mapping of accretion discs}
% allows explicit linebreak for the table of content
%
%
\titlerunning{Eclipse mapping of accretion discs}
% allows abbreviation of title, if the full title is too long
% to fit in the running head
%
\author{Raymundo Baptista\inst{1}}
\authorrunning{R. Baptista}
% if there are more than two authors,
% please abbreviate author list for running head
%
%
\institute{Departamento de F\'{\i}sica, UFSC, Campus Trindade, 
	88040-900, Florian\'opolis, Brazil}

\maketitle              % typesets the title of the contribution

\begin{abstract}
The eclipse mapping method is an inversion technique that makes use of 
the information contained in eclipse light curves to probe the structure, 
the spectrum and the time evolution of accretion discs. In this review
I present the basics of the method and discuss its different 
implementations. I summarize the most important results obtained to date 
and discuss how they have helped to improve our understanding of accretion
physics, from testing the theoretical radial brightness temperature 
distribution and measuring mass accretion rates to showing the evolution 
of the structure of a dwarf novae disc through its outburst cycle, from
isolating the spectrum of a disc wind to revealing the geometry of disc 
spiral shocks. I end with an outline of the future prospects.
\end{abstract}

\section{Introduction}

Accretion discs are cosmic devices that allow matter to efficiently 
accrete over a compact source by removing its angular momentum via 
viscous stresses while transforming the liberated gravitational
potential energy into heat and, thereafter, radiation \cite{acpower}.
They are widespread in astrophysical environments, from
sheltering the birth of stars to providing the energetics of quasars
and active galactic nuclei.
It is, however, in mass-exchanging binaries such as non-magnetic
Cataclysmic Variables (CVs) that the best environment for studies of 
accretion discs are possibly found. In these close binaries mass is fed
to a white dwarf by a Roche lobe filling companion star (the secondary)
via an accretion disc, which usually dominates the ultraviolet and 
optical light of the system \cite{bible}. 

Accretion discs in CVs cover
a wide range of accretion rates, \.{M}, and viscosity regimes.
For example, the sub-class of dwarf novae comprises low-mass transfer 
CVs showing recurrent outbursts (of 2--5 magnitudes, on timescales 
of weeks to months) which reflect changes in the structure of the discs 
-- from a cool, optically thin, low viscosity state to a hot, optically 
thick, high viscosity state -- and which are usually parameterized as a
large change in the mass accretion rate (\.{M}$= 10^{-11} \; M_\odot \;
yr^{-1} \mapsto 10^{-9} \; M_\odot \; yr^{-1}$) \cite{pvw}.
On the other hand, nova-like variables seem to be permanently in a high
viscosity state, presumably as a result of the fact that the accretion 
rate is always high.

The temperatures in CV discs may vary from 5000~K in the outer regions 
to over 50000~K close to the disc centre, and the surface density 
may vary by equally significant amounts over the disc surface. 
Therefore, the spectrum emitted by different
regions of the accretion disc may be very distinct.
Additionally, the bright spot (formed by the impact of the gas stream 
from the inner Lagrangian point on the disc rim), the white dwarf at 
disc centre, and the secondary star may all contribute to the integrated 
light of the binary.
Because what one directly observes is the combination of the spectra
emitted from these diverse regions and sources, the interpretation of 
disc observations is usually plagued by the ambiguity associated with 
composite spectra.
The most effective way to overcome these difficulties is with spatially
resolved data.

Two complementary indirect imaging techniques were developed in the 
1980's that provide spatially resolved observational constraints on 
accretion discs on angular scales of micro arcseconds -- well beyond 
the capabilities of current direct imaging techniques. One is Doppler 
Tomography \cite{mh88}, which is treated in detail elsewhere is this book
\cite{doppler}. It uses the changes in line emission profile with orbital
phase to probe the dynamics of accretion discs and is applicable to 
binaries over a large range of orbital inclinations, although it is
restricted to emission line data. 

The other is Eclipse Mapping \cite{h85}. 
It assembles the information contained in the shape of the eclipse
into a map of the accretion disc surface brightness distribution. 
While its application is restricted to deeply eclipsing binaries,
eclipse mapping can be used with continuum as well as line data.
When applied to time-resolved spectroscopy
through eclipses this technique delivers the spectrum of the disc at
any position on its surface.  Information on the radial dependence of
the temperature and vertical temperature gradients (for optically thick
regions), or temperature, surface density and optical depth (where the
disc is optically thin) can be obtained by comparing such spectra with
the predictions of models of the vertical disc structure.  The spatial
structure of the emission line regions over the disc can be similarly
mapped from data of high spectral resolution. Moreover, an eclipse map 
yields a snapshot of an accretion disc at a given time. By eclipse mapping
an accretion disc at different epochs it is possible to follow the secular 
changes of its radial brightness temperature distribution -- for example,
thought the outburst cycle of a dwarf nova -- allowing crucial tests of
accretion disc instability and viscosity models.

After more than a decade of experiments, eclipse mapping has now become
a mature and well established technique. There are already many good 
reviews on this topic in the literature \cite{hm86,h93,bible,w94}. 
The main aim of this review is therefore not to provide another 
description of the technique, but to make a summary of the results
obtained so far giving emphasis on the impact they had on our
understanding of the physics of accretion discs.

\section{The maximum entropy eclipse mapping method}

\subsection{The principles}

The three basic assumptions of the standard eclipse mapping method are:
(i) the surface of the secondary star is given by its Roche equipotential,
(ii) the brightness distribution is constrained to the orbital plane, and
(iii) the emitted radiation is independent of the orbital phase. 
The first assumption seems reasonably robust. The others are 
simplifications that do not hold in all situations. 
A discussion on the departures from assumptions (ii) and (iii) is 
presented in section~\ref{flare}.

A grid of intensities centred on the white dwarf, the eclipse map,
is defined in the orbital plane.
The eclipse geometry is specified by the inclination $i$, the binary mass
ratio $q$ (=$M_2/M_1$, where $M_2$ and $M_1$ are the masses of, 
respectively, the secondary star and the white dwarf) and the phase of 
inferior conjunction $\phi_0$ \cite{h85,h93}.
Given the geometry, a model eclipse light curve can be calculated for any 
assumed brightness distribution in the eclipse map. A computer code then
iteratively adjusts the intensities in the map (treated as independent
parameters) to find the brightness distribution the model light curve
of which fits the data eclipse light curve within the uncertainties. 
The quality of the fit is checked with a consistency statistics,
usually the reduced $\chi^2$. Because the one-dimensional data light 
curve cannot fully constrain a two-dimensional map, additional freedom
remains to optimize some map property. A maximum entropy (MEM) procedure
\cite{sb84,s87} is used to select, among all possible solutions, the one
that maximizes the entropy of the eclipse map with respect to a smooth 
default map.

\begin{figure}
\begin{center}
\includegraphics[bb=1.7cm 2.5cm 19.5cm 23.8cm,scale=0.7,clip]{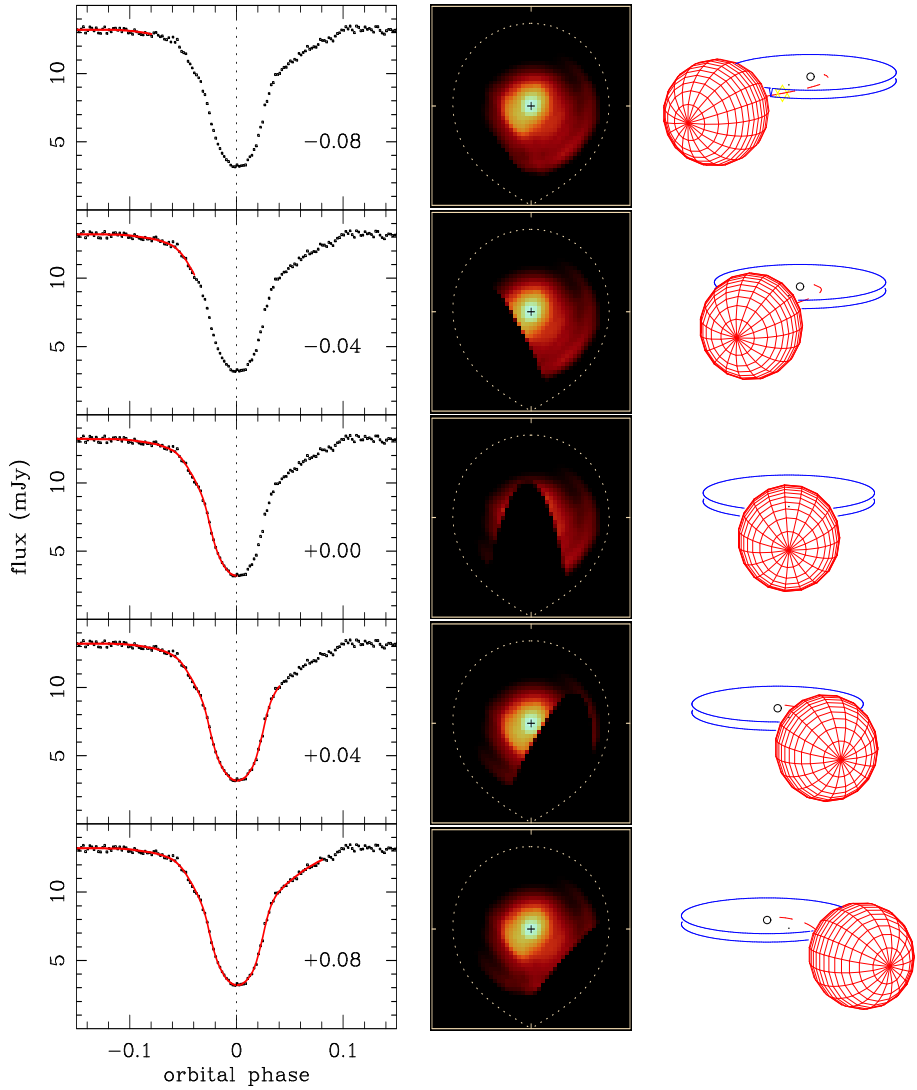}
\end{center}
\caption[]{ Simulation of a disc eclipse ($q=0.3, i=78\degr$).
  Left-hand panels: data light curve (dots) and model light curve
  (solid line) for five different orbital phases (indicated in the lower 
  right corner). Middle panels: eclipse maps in a false color blackbody
  logarithmic scale. Roche lobes for $q=0.3$ are shown as dotted lines; 
  crosses mark the centre of the disc. The secondary is below each panel 
  and the stars rotate counter-clockwise. Right-hand panels: the 
  corresponding geometry of the binary for each orbital phase.}
\label{demo}
\end{figure}

Figure~\ref{demo} illustrates the simulation of the eclipse of a fitted
brightness distribution while showing the comparison between the resulting 
model light curve and the data light curve.
The geometry in this case is $q=0.3$ and $i=81\degr$. 
%
%%%%%%%%%%%%%%%%%%%%%%%%%%%  FIGURE 1  %%%%%%%%%%%%%%%%%%%%%%%%%%%%%%
%%%%%%%%%%%%%%%%%%%%%%%%%%%%%%%%%%%%%%%%%%%%%%%%%%%%%%%%%%%%%%%%%%%%
%
The left-hand panels show the data light curve (small dots) and the model
light curve (solid line) as it is being drawn at five different orbital
phases along the eclipse (indicated in the lower right corner).
The right-hand panels depict the corresponding geometry of the binary for 
each orbital phase, in which the secondary star progressively occults the
accretion disc as well as the white dwarf and the bright spot.
The middle panels show the best-fit disc brightness distribution and how
it is progressively covered by the dark shadow of the secondary star 
during the eclipse. At phase $\phi=-0.08$ only a small fraction of the 
outer, faint disc regions are eclipsed and there is only a small reduction 
in flux in the light curve. The eclipse of the bright spot at the edge of
the disc and of the bright inner disc regions occur at about the same time
(slightly after $\phi=-0.04$) and coincide with the steepest ingress 
in the light curve. 
The flux at phase $\phi=0$ does not go to zero because a significant 
fraction of the disc remains visible at mid-eclipse. 
The asymmetry in the egress shoulder of the light curve maps into an
enhanced brightness emission in the trailing side of the disc (the 
right hemisphere of the eclipse map in Fig.~\ref{demo}).

\subsection{The expressions}

The expressions governing the eclipse mapping problem are as follows.
One usually adopts the distance from the disc centre to the internal 
Lagrangian point, $R_{L1}$, as the length scale. With this definition the
primary lobe has about the same size and form for any reasonable value 
of the mass ratio $q$ \cite{h85}.
If the eclipse map is an $N$ points flat, square grid of side 
$\lambda R_{L1}$, each of its surface element (pixel) has an area 
$(\lambda R_{L1})^2/N$ and an associated intensity $I_{j}$. 
The solid angle comprised by each pixel as seen from the earth is then
\begin{equation}
\theta^{2} = \left[ \frac{(\lambda R_{L1})^2}{N} \frac{1}{d^2} \right] 
\cos\,i \; ,
\end{equation}
where $d$ is the distance to the system. The value of $\lambda$ defines 
the area of the eclipse map while the choice of $N$ sets its spatial
resolution.
 
The entropy of the eclipse map $p$ with respect to the default map $q$ 
is defined as
\begin{equation}
S = - \sum_{j=1}^{N} \, p_j\,\ln \left( \frac{p_{j}}{q_{j}} \right) \; ,
\label{eq:1}
\end{equation}
where $p$ and $q$ are written as
\begin{equation}
p_{j} = \frac{I_{j}}{\sum_{k}I_{k}}
\;\;\;\;\; , \;\;\;\;\; q_{j} = \frac{D_{j}}{\sum_{k}D_{k}} \, .
\end{equation}
Other functional forms for the entropy appear in the literature 
\cite{h93,bible}. These are equivalent to (\ref{eq:1}) when $p$ and $q$ 
are written in terms of proportions.

The default map $D_j$ is generally defined as a weighted average of the
intensities in the eclipse map,
\begin{equation}
D_{j} = \frac{\sum_{k} \, \omega_{jk}I_{k}}{\sum_{k} \, \omega_{jk}}\; ,
\end{equation}
where the weight function $\omega_{jk}$ is specified by the user.
A priori information about the disc (p.ex., axi-symmetry) is included
in the default map via $\omega_{jk}$.
Prescriptions for the weight function $\omega_{jk}$ are discussed
in section \ref{defmap}.
In the absence of any constraints on $I_j$, the entropy has a maximum 
$S_{max}=0$ when $p_j=q_j$, or when the eclipse map and the default map 
are identical.

The model eclipse light curve $m(\phi)$ is derived from the intensities 
in the eclipse map,
\begin{equation}
m(\phi) = \theta^{2}\, \sum_{j=1}^{\rm N} \, I_{j}V_j(\phi) \; ,
\label{eq:2}
\end{equation}
where $\phi$ is the orbital phase.
The occultation function $V_j(\phi)$ specifies the fractional visibility 
of each pixel as a function of orbital phase and may include 
fore-shortening and limb darkening factors \cite{h93,r98,w94}.
The fractional visibility of a given pixel may be obtained by dividing the
pixel into smaller tiles and evaluating the Roche potential along the line
of sight for each tile to see if the potential falls below the value of the
equipotential that defines the Roche surface. If so, the tile is occulted. 
The fractional visibility of the pixel is then the sum of the visible tiles
divided by the number of tiles.

The consistency of an eclipse map may be checked using the $\chi^2$ as a
constraint function,
\begin{equation}
\chi^{2} = \frac{1}{M}\:\sum_{\phi=1}^{M} \, 
\left( \frac{m(\phi)-d(\phi)}{\sigma(\phi)} \right)^2 =
\frac{1}{M}\:\sum_{\phi=1}^{M} \, r({\phi})^{2}\; ,
\label{eq:chi}
\end{equation}
where $d(\phi)$ is the data light curve, $\sigma(\phi)$ are the 
corresponding uncertainties, $r(\phi)$ is the residual at the orbital
phase $\phi$, and $M$ is the number of data points in the light curve.
Alternatively, the constraint function may be a combination of the
$\chi^{2}$ and the R-statistics \cite{bs93},
\begin{equation}
R = \frac{1}{\sqrt{M-1}}\:\sum_{\phi=1}^{M-1}\,r({\phi}) \;
 r({\phi+1})\; ,
\label{eq:r}
\end{equation}
to minimize the presence of correlated residuals in the model light 
curve \cite{bs91}. For the case of uncorrelated normally distributed
residuals, the R-statistics has a Gaussian probability distribution 
function with average zero and unity standard deviation. Requiring the 
code to achieve an $R=0$, is equivalent to asking for a solution with 
uncorrelated residuals in the model light curve.

The final MEM solution is the eclipse map that is as close
as possible to its default map as allowed by the constraint imposed by
the light curve and its associated uncertainties \cite{h93,r98}.
In matematical terms, the problem is one of constrained maximization, 
where the function to maximize is the entropy and the constraint is a
consistency statistics that measures the quality of the fitted model to 
the data light curve.
Different codes exist to solve this problem. Many of the eclipse mapping 
codes are based on the commercial optimization package MEMSYS \cite{sb84}.
Alternative implementations using conjugate-gradients algorithms
\cite{bs91,bs93}, CLEAN-like algorithms \cite{spruit94} and, more recently, genetic algorithms \cite{bob00} are also being used.

\subsection {Default maps} \label{defmap}

A crucial aspects of eclipse mapping is the selection of the weight 
function for the default map, which allows the investigator to steers 
the MEM solution towards a determined type of disc map. A list of
different prescriptions for $\omega_{jk}$ is given in Table~\ref{tab1}.
%
%%%%%%%%%%%%%%%%%%%%%%%%  TABLE 1  %%%%%%%%%%%%%%%%%%%%%%%%%%%%
\begin{table}
\caption{Prescriptions for default maps}
\begin{center}
\renewcommand{\arraystretch}{1}
\setlength\tabcolsep{6pt}
\begin{tabular}{llc}
\hline \\ [-1ex]
\multicolumn{2}{c}{weight functions} & reference \\ [1ex]
\hline \\ [-1ex]
A) most uniform map: & $\omega_{jk} = 1$ & \cite{h85} \\ [3ex]

B) smoothest map: & $\omega_{jk} = \exp \left( - \frac{d_{jk}^2}
{2 \;\Delta^2} \right)$ & \cite{h85} \\ [3ex]

C) most axisymmetric map: & $\omega_{jk} = \exp \left[ - \frac{(R_j - R_k)^2}
{2\;\Delta^2} \right]$ & \cite{h85} \\ [-1ex]
~~~(full azimuthal smearing) \\ [3ex]

D) limited azimuthal smearing: & $\omega_{jk} = \exp \left[ - \frac{1}{2}
\left( \frac{(R_j - R_k)^2}{\Delta_R^2} + \frac{\theta_{jk}^2}
{\Delta_\theta^2} \right) \right]$ & \cite{r93} \\ [-1ex]
~~~(constant angle $\theta$) \\ [3ex]

E) limited azimuthal smearing: & $\omega_{jk} = \exp \left[ - \frac{1}{2}
\left( \frac{(R_j - R_k)^2}{\Delta_R^2} + \frac{s_{jk}^2}{\Delta_s^2} \right)
\right]$ & \cite{bsh} \\ [-1ex]
~~~(constant arc length $s$) \\ [2ex]
\hline \\ [-2ex]
\multicolumn{3}{l}{NOTES: $d_{jk}$ is the distance between pixels $j$ and $k$;
$R_j$ and $R_k$ are the distances} \\
\multicolumn{3}{l}{ from pixels $j$ and $k$ to the centre of the disc; 
$\theta_{jk}$ is the azimuthal angle between} \\
\multicolumn{3}{l}{ pixels $j$ and $k$; and $s_{jk}$ is the arc-length 
between pixels $j$ and $k$.} \\ [-3ex]
\end{tabular}
\end{center}
\label{tab1}
\end{table}
%%%%%%%%%%%%%%%%%%%%%%%%%%%%%%%%%%%%%%%%%%%%%%%%%%%%%%%%%%%%%%%

Choosing $\omega_{jk}=1$ (option A) results in a uniform default map and 
will lead to the {\em most uniform eclipse map} consistent with the data. 
This happens not to be a good choice for eclipse mapping because it
results in a map severely distorted by criss-crossed artifacts 
\cite{bob97,h85,spruit94}. 
This effect may be reduced by setting the weight function as a Gaussian 
profile of width $\Delta$ (option B), which results in the {\em smoothest
map} that fits the data.

The third case (option C) sets $D_j$ as an axi-symmetric average of the 
eclipse map and will lead to the {\em most nearly axi-symmetric map} that 
fits the data. 
It suppresses the azimuthal information in the default map while keeping 
the radial structure of $I_{j}$ on scales greater than $\Delta_R$. This
seems a reasonable choice for accretion disc mapping because one expects
the disc material to be roughly in Keplerian orbits, so that local 
departures from axi-symmetry will tend to be diminished by the strong shear.
This is a commonly used option and is also known as the default map of full
azimuthal smearing.

The full azimuthal smearing default results in rather distorted
reproduction of asymmetric structures such as a bright spot at the disc
rim. In this case, the reconstructed map exhibits a lower integrated
flux in the asymmetric source region, the excess being redistributed as
a concentric annulus about the same radial distance.
By limiting the amount of azimuthal smearing it is possible to alleviate
this effect and to start recovering azimuthal information in the accretion
disc. Two prescriptions in this regard were proposed. Reference \cite{r93}
limited the amount of azimuthal smearing by averaging over a polar Gaussian 
weight function of {\em constant angles} along the map (option D) while
\cite{bsh} chosen to use a polar Gaussian function of {\em constant arc 
length} through the map (option E). 

\begin{figure}[t]
\begin{center}
\includegraphics[bb=6cm 6.8cm 16cm 21.8cm,scale=.5]{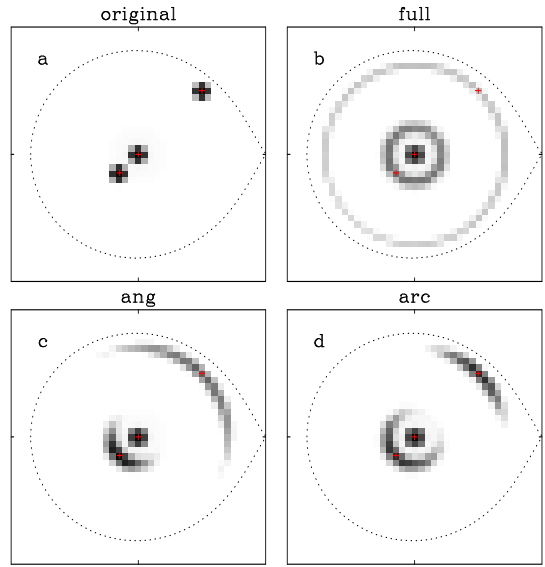}
\end{center}
\caption[]{Effects of the different weight functions for the default map. 
 % Maps are shown in a logarithmic greyscale; dark regions are brighter. 
  (a) The original map, with three Gaussian spots. The corresponding default 
  map obtained using the default of (b) full azimuthal smearing; (c) 
  constant angles; and (d) constant arc length. From \cite{bsh}. }
\label{fdef}
\end{figure}

Figure~\ref{fdef} shows the effects of the three last weight functions 
when applied to an artificial map containing three Gaussian spots
at different radial distances from its centre. 
%
%%%%%%%%%%%%%%%%%%%%%%%%%%%  FIGURE 2  %%%%%%%%%%%%%%%%%%%%%%%%%%%%%%
%%%%%%%%%%%%%%%%%%%%%%%%%%%%%%%%%%%%%%%%%%%%%%%%%%%%%%%%%%%%%%%%%%%%
The default with
constant angles is more efficient to reproduce asymmetries in the inner
disc regions (such as an accretion column, or the expected dipole
pattern for velocity-resolved line emission mapping, or in presence of
a bipolar wind emanating from the inner disc), while the default of
constant arc length is more efficient in recovering asymmetries in the
outer parts of the disc (such as a bright spot at disc rim, or in the
case of an eccentric disc).  The choice between these two default
functions, in a given case, is defined by whether it is more important
to have a better azimuthal resolution at the inner or outer disc
regions. The radial profile is not affected by this choice.

Other possibilities concerning the default map were proposed in terms of
the combination of different weight functions \cite{bob97,spruit94}.
Particularly, the mix of the smoothest and the most axi-symmetric defaults
is, in a sense, equivalent to the default of limited azimuthal smearing 
and leads to similar results.
Another interesting proposal is that of a negative weight function, 
that may be used to avoid or minimize a certain map property (p.ex., 
the presence of the undesired criss-crossed arcs) \cite{spruit94}.

\subsection {The uneclipsed component} \label{fbg}

Reference \cite{rpt92} found that the entropy function can be a useful
tool to signal and to isolate the fraction of the total light which is
not coming from the accretion disc plane.  They noted that when the
light curve is contaminated by the presence of additional light (e.g.,
from the secondary star) the reconstructed map shows a spurious structure 
in the regions farthest away from the secondary star (the upper lune of
the eclipse maps in Fig.~\ref{demo}, hereafter called the `back' side of
the disc).  This is because the eclipse mapping
method assumes that all the light is coming from the accretion disc, in
which case the eclipse depth and width are correlated in the sense that
a steeper shape corresponds to a deeper eclipse.  The addition of an
uneclipsed component in the light curve (i.e., light from a source
other than the accretion disc) ruins this correlation.  To account for
the extra amount of light at mid-eclipse and to preserve the brightness
distribution derived from the eclipse shape the algorithm inserts the
additional light in the region of the map which is least affected by
the eclipse, leading to a spurious front-back disc brightness asymmetry.
Since the entropy measures the amount of structure in the map, the 
presence of these spurious structures is flagged with lower
entropy values.  

The correct offset level may be found by comparing a
set of maps obtained with different offsets and selecting the one with
highest entropy.  Alternatively, the value of the zero-intensity level
can be included in the mapping algorithm as an additional free
parameter to be fitted along with the intensity map in the search for
the MEM solution \cite{b95,r94}. A detailed discussion on the 
reliability and consistency of the estimation of the uneclipsed
component can be found in \cite{bsh}.

\subsection {Beyond the standard assumptions} \label{flare}

The standard eclipse mapping assumes a simple flat, geometrically thin
disc model. Real discs may however violate this assumption in the limit
of high \.{M}. Disc opening angles of $\alpha \simgt 4 \degr$ are predicted 
for \.{M}$\simgt 5 \times 10^{-9}\;M_\odot\,yr^{-1}$ \cite{mm82,smak92}.
At large inclinations ($i\simgt 80 \degr$) this may lead to artificial
front-back asymmetries in the eclipse map (similar to those discussed in section~\ref{fbg}) because of the different effective areas of surface 
elements in the front and back sides of a flared disc as seen by an 
observer on Earth. In extreme cases, this may lead to obscuration of 
the inner disc regions by the thick disc rim (e.g., \cite{knigge00}). 

Motivated by the front-back asymmetry that appeared in the flat-disc
map and by the difficulties in removing the asymmetry with the assumption
of an uneclipsed component, \cite{rob95,rob99} introduced a flared disc in
their eclipse mapping of ultraviolet light curves of the dwarf nova Z~Cha
at outburst maximum. They found that the asymmetry vanishes and the
disc is mostly axi-symmetric for a disc opening angle of $\alpha= 6\degr$.

\begin{figure}[t]
\begin{center}
%\vspace*{6.5cm}
\includegraphics[clip,bb=1.8cm 9.8cm 20cm 19.7cm,scale=.7]{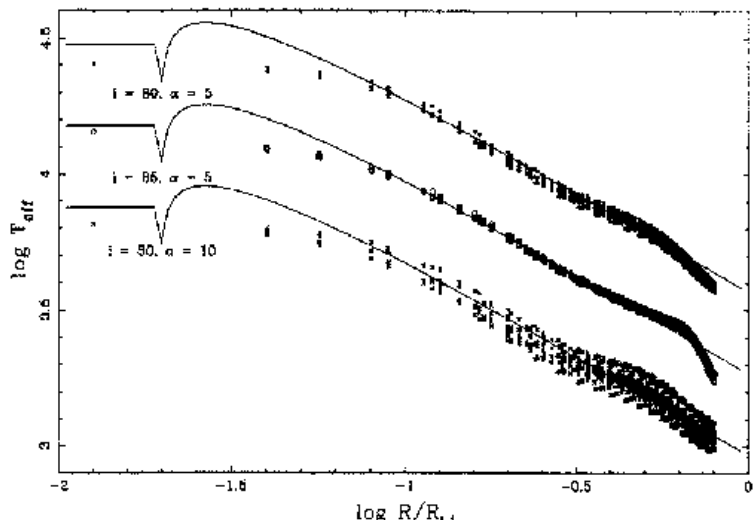}
\end{center}
\caption[]{Examples of reconstructed radial brightness temperature
  distributions with the flat-disc eclipse mapping for the case of
  flared discs. The inclinations $i$ and disc opening angle $\alpha$ 
  are indicated for each case. From \cite{r98}.}
\label{fig3}
\end{figure}

Simulations \cite{r98} show that eclipse mapping reconstructions obtained 
with the flat-disc assumption result in good reproduction of the radial
temperature distribution of flared accretion discs provided that the inner
disc regions are not obscured by the disc rim (Fig.~\ref{fig3}).
%
%%%%%%%%%%%%%%%%%%%%%%%%%%%  FIGURE 3  %%%%%%%%%%%%%%%%%%%%%%%%%%%%%%
%%%%%%%%%%%%%%%%%%%%%%%%%%%%%%%%%%%%%%%%%%%%%%%%%%%%%%%%%%%%%%%%%%%%

Reference \cite{w94} shows that it is usually impossible to distinguish 
between a flared disc and an uneclipsed component to the total light. 
Both effects lead to the appearance of spurious structures in the
back regions of the disc, and eclipse maps obtained with either model 
may lead to equally good fits to the data light curve. 
Reference \cite{bc00a} pointed out that spectral eclipse mapping could 
help in evaluating the importance of each of these effects in a given case.
If the uneclipsed component is caused by an optically-thin,
vertically-extended disc wind, the uneclipsed spectrum shows a Balmer 
jump in emission plus strong emission lines, while in the case of a 
flared disc the spurious uneclipsed spectrum should reflect the difference
between the disc spectrum of the back (deeper atmospheric layers seen 
at lower effective inclinations) and the front (upper atmospheric 
layers seen at grazing incidence) sides and should mainly consist 
of continuum emission filled with absorption lines.

Because of the assumption that the emitted radiation is independent of 
the orbital phase, in the standard eclipse mapping method all variations
in the eclipse light curve are interpreted as being caused by the 
changing occultation of the emitting region by the secondary star.
Thus, out-of-eclipse brightness changes (p.ex., orbital modulation
due to anisotropic emission from the bright spot)
has to be removed before the light curves can be analyzed.
The usual approach is to interpolate the out-of-eclipse light curve 
across the eclipse phases \cite{h85,rpt92}. An alternative approach
is to apply a light curve decomposition technique to separate the
contributions of the white dwarf, bright spot and accretion disc
\cite{wood86,wood89}. This technique however requires high signal-to-noise
light curves and good knowledge of the contact phases of the white
dwarf and bright spot, which limits its application to a few objects.

A step to overcome these limitations was done by \cite{bob97}
with the inclusion of a disc rim in the eclipse mapping method.
The out-of-eclipse modulation is modeled as the fore-shortening of
an azimuthally-dependent brightness distribution in the disc rim. 
This procedure allows to recover the azimuthal (phase) dependency of 
the bright spot emission. It however requires a good estimate of the 
outer disc radius.

The more advanced code of \cite{r98}, including a flared disc, the disc 
rim, and the surface of the Roche-lobe filling secondary star, expanded 
the eclipse mapping method into a three-dimensional mapping technique.
Nevertheless, it comes along with a significant increase in the degrees 
of freedom that aggravates the problem of non-uniqueness of solutions.

\subsection{Performance under extreme conditions} \label{tests}

In an eclipse mapping reconstruction,
the brightness of a given surface element is derived from the 
information given by the changes in flux caused by its occultation 
(at ingress) and reappearance (at egress) from behind the secondary star.
In the case of an eclipse light curve with incomplete phase coverage,
there are regions in the disc for which only one of these pieces of 
information is available. Moreover, for a system with low inclination,
there are regions in the back side of the disc which are never covered
by the shadow of the secondary star and, therefore, there is no 
information about the brightness distribution of these regions on the
shape of the light curve.
This section presents simulations aiming to assess the reliability of
eclipse mapping reconstructions obtained under the combined extreme 
conditions of incomplete eclipse coverage, low binary inclination and 
relatively low signal-to-noise data.

\begin{figure}[t]
\begin{center}
\includegraphics[bb=1.4cm 8.5cm 19cm 20cm,angle=-90,scale=.5]{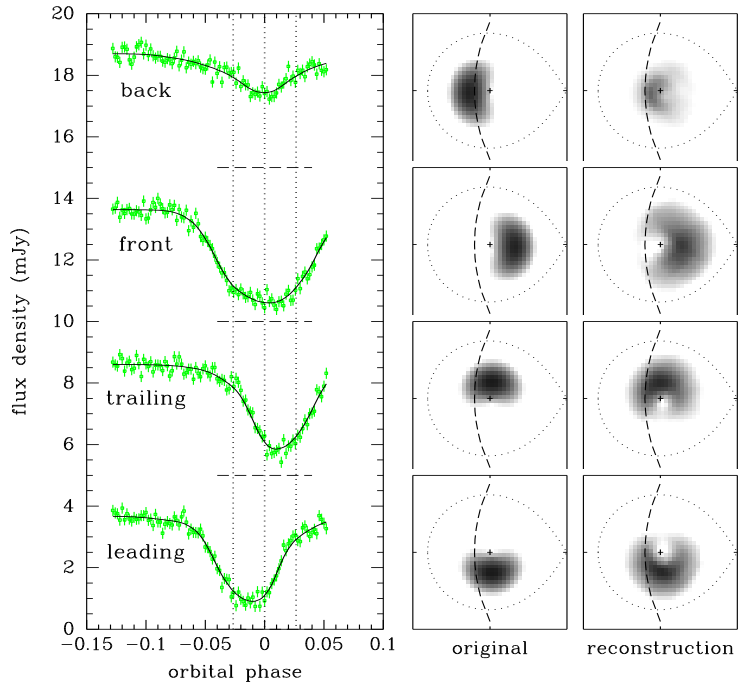}
\end{center}
\caption[]{Reconstructing asymmetric brightness distributions with light 
   curves of low signal-to-noise and incomplete phase coverage. The
   left-hand panel shows the artificial light curves (dots with error bars)
   and corresponding eclipse mapping models (solid lines). Horizontal 
   dashed lines indicate the true zero level in each case. Vertical dotted
   lines mark ingress/egress phases of the white dwarf and mid-eclipse. 
   The middle and right-hand panels show, respectively, the original maps 
   and the reconstructions in a logarithmic greyscale. 
   Bright regions are dark; faint regions are white. A cross marks the 
   center of the disc; dotted lines show the Roche lobe and dashed lines 
   depict the locus of the far edge of the shadow of the secondary star 
   along the eclipse. The secondary is to the right of each map and the 
   stars rotate counter-clockwise. }
\label{simul}
\end{figure}

Four artificial brightness distributions with asymmetric polar Gaussian 
spots on the back, front, leading and trailing sides of the disc were
constructed (Fig.~\ref{simul}). A low-inclination geometry ($q=1$ and
$i=71\degr$) was adopted to simulate the eclipses and light curves with
signal-to-noise $S/N\simeq 5-15$ and an incomplete set of orbital
phases were produced.
The artificial light curves were analyzed with the eclipse mapping method
and the results are shown in Fig.~\ref{simul}.
%
%%%%%%%%%%%%%%%%%%%%%%%%%%%  FIGURE 4  %%%%%%%%%%%%%%%%%%%%%%%%%%%%%%
%%%%%%%%%%%%%%%%%%%%%%%%%%%%%%%%%%%%%%%%%%%%%%%%%%%%%%%%%%%%%%%%%%%%  

For the adopted set of orbital phases, the leading side of the disc 
(the lower hemisphere of the eclipse maps in Fig.~\ref{simul}) is 
mapped by the moving shadow of the secondary star both during ingress
and egress, whereas much of the trailing side of the disc is only mapped
by the secondary star at ingress phases.
Dashed lines in the eclipse maps mark the locus of the far edge of the 
shadow of the secondary star along the eclipse. Regions to the left of 
this line are never covered by the secondary star.

Despite the incomplete eclipse coverage, good quality reconstructions 
are obtained for the front, trailing and leading maps. The spots
appear spread in radius due to the low signal-to-noise of the light 
curves, and are elongated in azimuth because of the intrinsic azimuthal 
smearing effect of the eclipse mapping method.
The results are equally good for the leading and the trailing maps,
despite the fact that the spot in the latter case is located in the 
disc region for which there is limited information in the shape of the 
light curve. 
For the back map, much of the asymmetric brightness distribution completely
escapes eclipse. Not surprisingly, the eclipse map does not correctly 
reproduce the brightness distribution in the disc regions beyond those 
covered by the secondary star. The missing flux appears in the uneclipsed
component.

These simulations show that eclipse mapping obviously fails to
recover the brightness distribution of disc regions for which there is
no information in the shape of the eclipse, but performs reasonably well
in the case of incomplete phase coverage even with relatively low
signal-to-noise data.

The brightness distribution of the back map approximately simulates the
intrinsic front-back asymmetry of a flared disc as seen at a high 
inclination angle ($i> 80\degr$). It is fortunately that, in these cases, 
the shadow of the secondary star maps most (if not all) of the primary 
Roche lobe for any reasonable mass ratio.

Simulations of reconstructions from light curves of more limited phase 
coverage are presented in \cite{bc00a}.
Tests on the reliability of eclipse mapping reconstructions under 
a variety of other conditions can be found in the literature
\cite{bhs,bsh,bob97,b3,h85,r98,spruit94}.

\section{A summary of results}

When eclipse mapping appeared in the mid-1980's, the standard picture of 
an accretion disc in a CV was that of a flat, nearly axi-symmetric disc 
with a bright spot on its edge. This section reviews some of the eclipse 
mapping results that helped to improve this picture either by allowing key 
tests of theoretical expectations or by revealing new and unexpected 
aspects of the physics of accretion discs.

\subsection{Classical results}

\begin{figure}
\includegraphics[bb=-5cm 6cm 12cm 24cm,scale=.4]{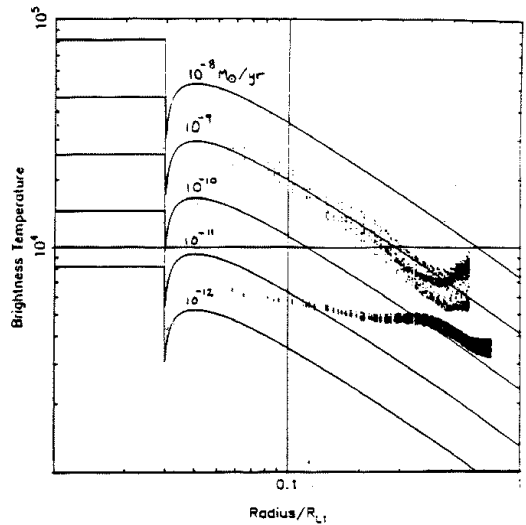}
\caption{ The radial temperature profile of the dwarf nova Z~Cha in 
  outburst and in quiescence. The solid lines show steady-state 
  blackbody disc models for mass accretion rates from $10^{-8}$ to 
  $10^{-12}\; M_\odot$\,yr$^{-1}$. From \cite{h93}. }
\label{fig5}
\end{figure}

Early applications of the technique were useful to show that accretion 
discs in outbursting dwarf novae \cite{hc85} and in long-period novalike
variables \cite{hs85,rpt92} closely follow the expected radial dependence
of temperature with radius for a steady-state disc, $T \propto R^{-3/4}$, 
and to reveal that the radial temperature profile is essentially flat 
in the short period quiescent dwarf novae \cite{wood86,wood89,wood92}
(Fig.~\ref{fig5}). This suggests that the viscosity in these short
period systems is much lower in quiescence than in outburst, lending 
support to the disc instability model, and that their quiescent discs
are far from being in a steady-state. 
Eclipse mapping studies also contributed to the puzzle about the
SW~Sex stars -- a group of mostly eclipsing novalike variables with 
periods in the range 3-4~hr that display a number of unexplained 
phenomena -- by showing that the radial temperature profile in these 
systems is noticeably flatter than the $T \propto R^{-3/4}$ law
\cite{bsh,rpt92}.
A flat radial temperature distribution is also suggested for the old
novae V~Per, that lies in the middle of the CV period gap \cite{was92}.
Unfortunately, this result is rather uncertain because it is based on
uncalibrated white-light data and there is a large uncertainty on the
eclipse geometry.
A comprehensive account of these pioneering results can be found in 
\cite{h93}.
%
%%%%%%%%%%%%%%%%%%%%%%%%%%%  FIGURE 5  %%%%%%%%%%%%%%%%%%%%%%%%%%%%%%
%%%%%%%%%%%%%%%%%%%%%%%%%%%%%%%%%%%%%%%%%%%%%%%%%%%%%%%%%%%%%%%%%%%%
%

It has been a usual practice to convert the intensities in the eclipse 
maps to blackbody brightness temperatures and then compare them to the
radial run of the effective temperature predicted by steady state, 
optically thick disc models. A criticism about this procedure is that a
monochromatic blackbody brightness temperature may not always be a proper
estimate of the disc effective temperature. As pointed out by \cite{b98}, 
a relation between these two quantities is non-trivial, and can only be 
properly obtained by constructing self-consistent models of the vertical
structure of the disc. Nevertheless, the brightness temperature should be
close to the effective temperature for the optically thick disc regions.

\begin{figure}
\includegraphics[bb=1.5cm -3.5cm 20cm 16cm,angle=-90,scale=.35]{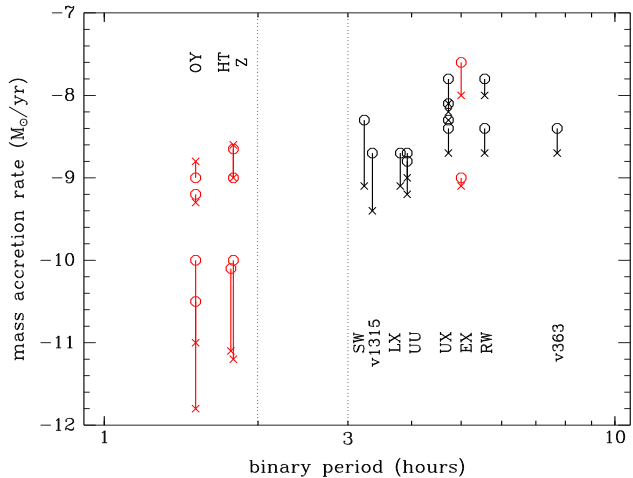}
\caption{ Mass transfer rates at radii of $0.1\;R_{L1}$ (crosses) 
and $0.3\;R_{L1}$ (circles) as a function of the binary period. }
\label{mdot}
\end{figure}

From the $T(R)$ diagram it is possible to obtain an independent estimate
of the disc mass accretion rate. Reference \cite{h93} compiled the
inferred mass accretion rates from a dozen of eclipse mapping 
experiments to construct an \.{M}$\times P_{orb}$ diagram. 
An updated version of this diagram is shown in Fig.~\ref{mdot}.
It seems a bit disappointing that the diagram is still loosely populated.
In particular, there is yet no eclipse mapping estimate of \.{M} for
a system inside the 2-3~hr CV period gap.
There is a significant scatter in the \.{M} derived from different 
experiments for a given object (e.g., in UX~UMa, from $10^{-8.1}$ to 
$10^{-8.7}\; M_\odot$\,yr$^{-1}$ at $0.1\;R_{L1}$). Whereas part of 
this scatter is possibly a real effect due to long-term changes in the 
mass transfer rates, it stands as a warning that one should be careful 
in interpreting mass accretion rates derived from the brightness 
temperature distributions, as discussed above.
%
%%%%%%%%%%%%%%%%%%%%%%%%%%%  FIGURE 6  %%%%%%%%%%%%%%%%%%%%%%%%%%%%%%
%%%%%%%%%%%%%%%%%%%%%%%%%%%%%%%%%%%%%%%%%%%%%%%%%%%%%%%%%%%%%%%%%%%%
%

According to current evolutionary scenarios, CVs should evolve towards 
shor\-ter orbital periods with decreasing mass transfer rates as a 
consequence of orbital angular momentum losses due mainly to magnetic 
braking (for systems above the period gap) or gravitational radiation 
(for systems below the gap) \cite{ps81,p84}.
In Fig.~\ref{mdot} it appears that there is a tendency among the 
steady-state discs of novalike variables to show larger \.{M} for longer 
binary period -- in agreement with the above expectation -- and that the 
discs of novalike variables and outbursting dwarf novae have comparable
\.{M}. The mass accretion rates in the eclipse maps of novalike variables 
increase with disc radius. The departures from the steady-state disc 
model are more pronounced for the SW~Sex stars (period range 3-4~hs).
Illumination of the outer disc regions by the inner disc or mass ejection 
in a wind from the inner disc are possible explanations for this effect.

\begin{figure}
\scalebox{.5}{%
\includegraphics[bb=1.3cm 6.2cm 20cm 24.5cm,clip,scale=.65]
{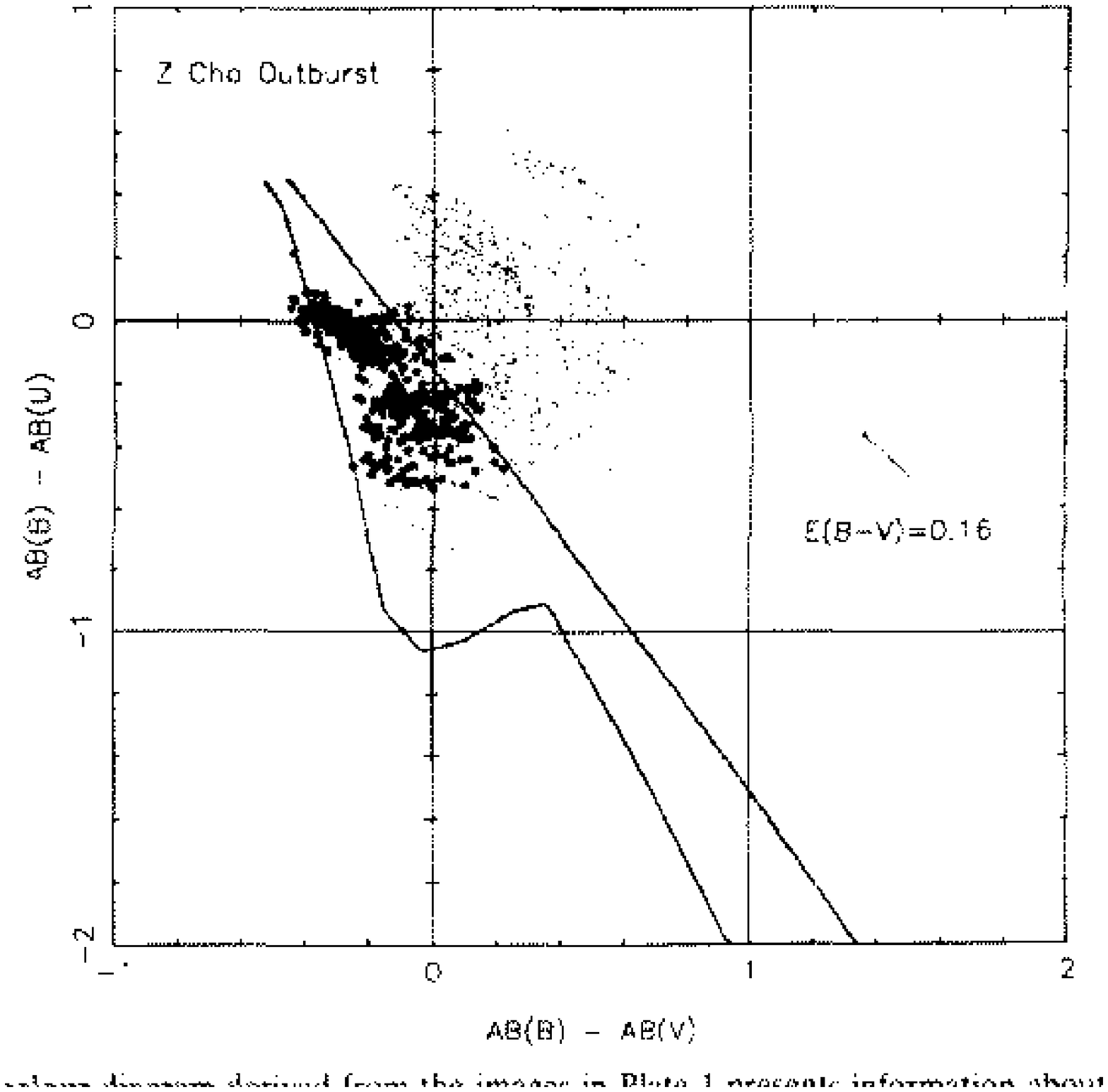} }%
\quad
\scalebox{.5}[.6]{%
\includegraphics[bb=2cm 7.5cm 19.5cm 21.8cm,clip,scale=.65]
{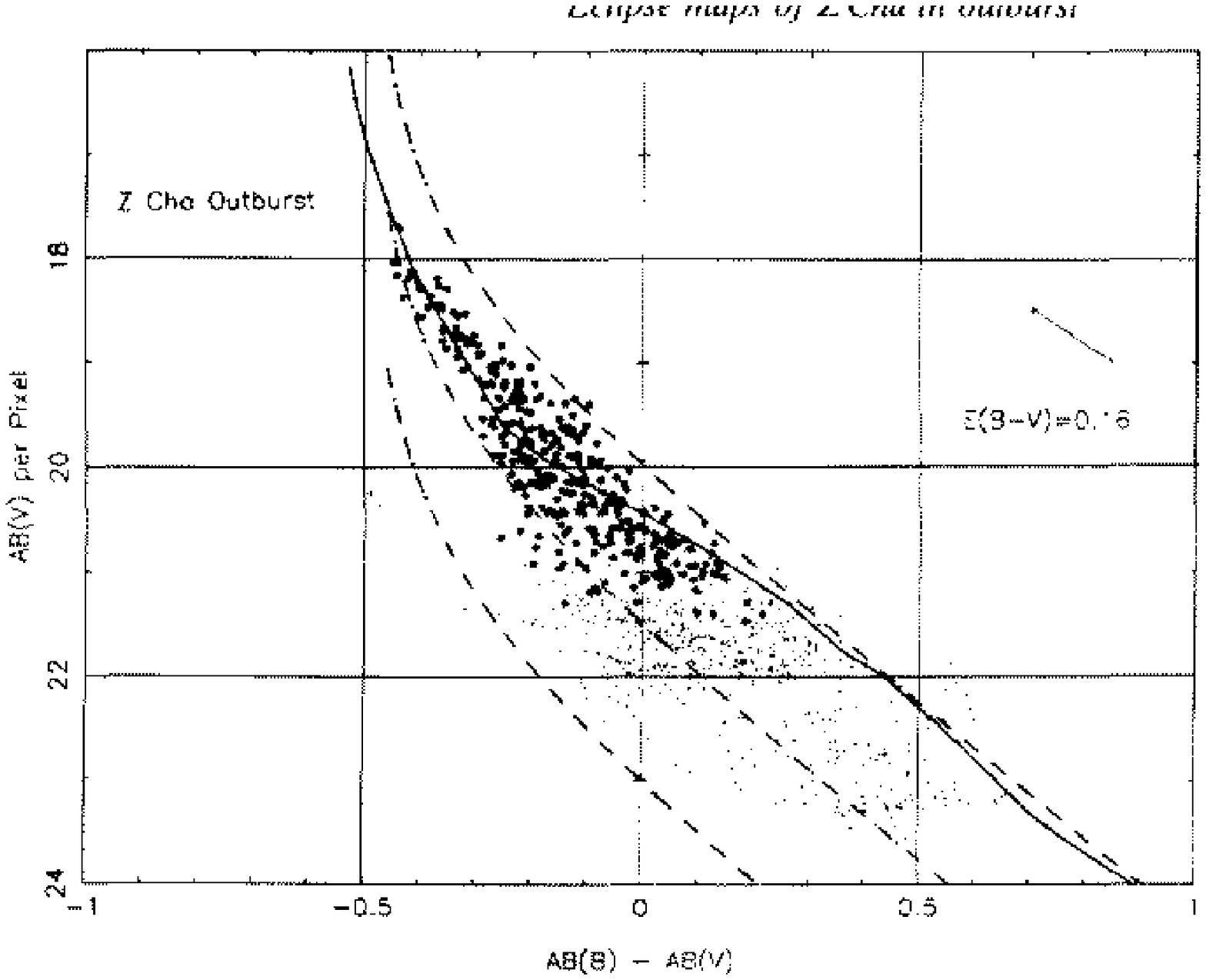} }
\caption{ Inferring the disc emission properties and the distance to Z~Cha
  from the two-colour and colour-magnitude diagrams. Surface elements at 
  the inner disc ($R< 0.3\;R_{L1}$) are represented by large dots, while 
  elements in the outer disc regions are indicated by small dots. The
  solid and dashed curves in the right-hand panel show, respectively, 
  the main sequence relationship for the best-fit distance and blackbody
  relationships for three different assumed distances.
  From \cite{hc85}. }
\label{fig7}
\end{figure}

Multi-colour eclipse mapping is useful to probe the spectrum emitted by 
the different parts of the disc surface. Two-colour diagrams show that 
the inner disc regions of outbursting dwarf novae \cite{b3,hc85} and of 
novalike variables \cite{b95,bsh,hs85} are optically thick with a vertical
temperature gradient less steep than that of a stellar atmosphere, and that
optically thin, chromospheric emission appears to be important in the 
outer disc regions (Fig.~\ref{fig7}).
The fact that the emission from the inner disc regions is optically thick
thermal radiation opens the possibility to use a colour-magnitude diagram
to obtain independent estimates of the distance to the binary with a 
procedure similar to cluster main-sequence fitting.
Distance estimates with this method were obtained for Z~Cha 
\cite{hc85}, OY~Car \cite{b3}, UU~Aqr \cite{bsh}, RW~Tri \cite{hs85} 
and UX~UMa \cite{b95}.
%
%%%%%%%%%%%%%%%%%%%%%%%%%%%  FIGURE 7  %%%%%%%%%%%%%%%%%%%%%%%%%%%%%%
%%%%%%%%%%%%%%%%%%%%%%%%%%%%%%%%%%%%%%%%%%%%%%%%%%%%%%%%%%%%%%%%%%%%
%

\subsection{Spectral studies}

The eclipse mapping method advanced to the stage of delivering
spatially-resolved spectra of accretion discs with its application to
time-resolved eclipse spectrophotometry \cite{r93}.
The time-series of spectra is divided up into numerous spectral bins 
and light curves are extracted for each bin. The light curves are then
analyzed to produce a series of monochromatic eclipse maps covering the 
whole spectrum.  Finally, the maps are combined to obtain the spectrum 
for any region of interest on the disc surface.

\begin{figure}
\includegraphics[bb=2.2cm 2.5cm 19.5cm 20cm,angle=-90,scale=.54]{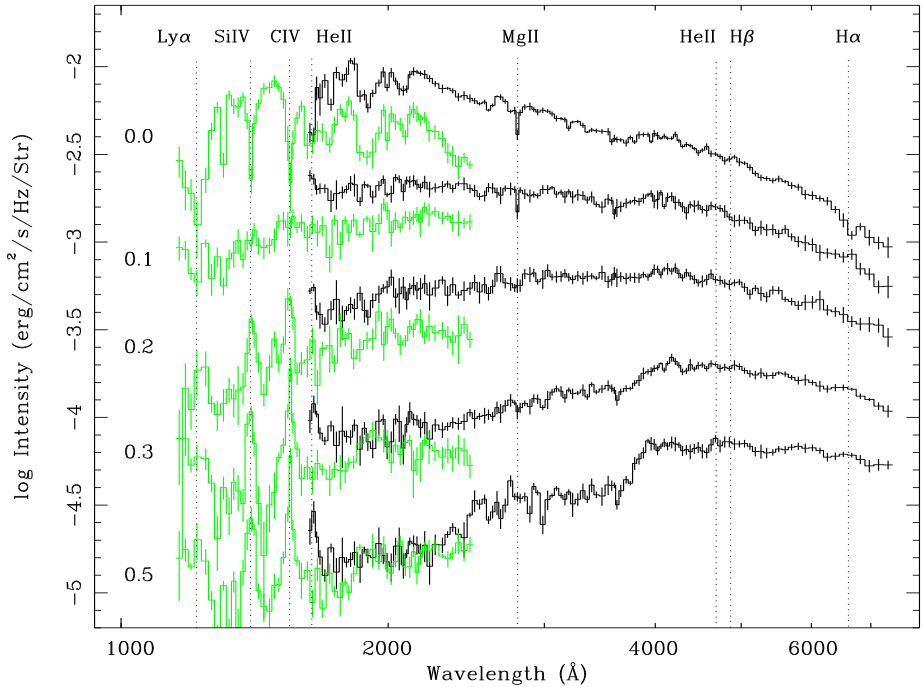}
\caption{ Spatially resolved spectra of the UX\,UMa accretion disc on
   August 1994 (gray) and November 1994 (black). The spectra were computed
   for a set of concentric annular sections (mean radius indicated on 
   the left, in units of $R_{L1}$). The most prominent line transitions 
   are indicated by vertical dotted lines. From \cite{b98}. }
\label{fig8}
\end{figure}

The spectral mapping analysis of the nova-like variables UX~UMa 
\cite{b95,b98,r93,r94} and UU~Aqr \cite{bssh} shows that the inner 
accretion disc is characterized by a blue continuum filled with 
absorption bands and lines which cross over to emission with increasing 
disc radius (Fig.~\ref{fig8}). The continuum emission becomes 
progressively fainter and redder as one moves outwards, reflecting 
the radial temperature gradient.
Similar results were found for SW~Sex \cite{gthesis} and RW~Tri \cite{groot}.
Not surprisingly, these high-\.{M} discs seem hot and optically thick in 
their inner regions and cool and optically thin in their outer parts.
%
%%%%%%%%%%%%%%%%%%%%%%%%%%%  FIGURE 8  %%%%%%%%%%%%%%%%%%%%%%%%%%%%%%
%%%%%%%%%%%%%%%%%%%%%%%%%%%%%%%%%%%%%%%%%%%%%%%%%%%%%%%%%%%%%%%%%%%%
%

However, the unprecedent combination of spatial and spectral resolution 
obtained with spectral mapping started to reveal a multitude of unexpected
details.
In UU Aqr, the lines show clear P~Cygni profiles at intermediate and 
large disc radii in an evidence of gas outflow \cite{bssh}.
In UX~UMa, the comparison of spatially resolved spectra at different 
azimuths reveals a significant asymmetry in the disc emission at ultraviolet
wavelengths, with the disc side closest to the secondary star showing 
pronounced absorption bands and a Balmer jump in absorption. 
This effect is reminiscent of that observed previously in OY Car, where 
the white dwarf emission seems veiled by an ``iron curtain'' \cite{h94}, 
and was attributed to absorption by cool circumstellar material \cite{b98}. 
The spectrum of the infalling gas stream in UX~UMa and UU~Aqr is 
noticeably different from the disc spectrum at the same radius suggesting 
the existence of gas stream ``disk-skimming'' overflow that can be seen
down to $R\simeq (0.1-0.2)\; R_{L1}$. Spectra at the site of the bright 
spot suggest optically thick gas, with the Balmer jump and the Balmer 
lines in absorption.

\begin{figure}
%\vspace*{7cm}
\scalebox{.5}[.61]{%
\includegraphics[bb=1cm 21.1cm 18.8cm 21.6cm,scale=.65]
{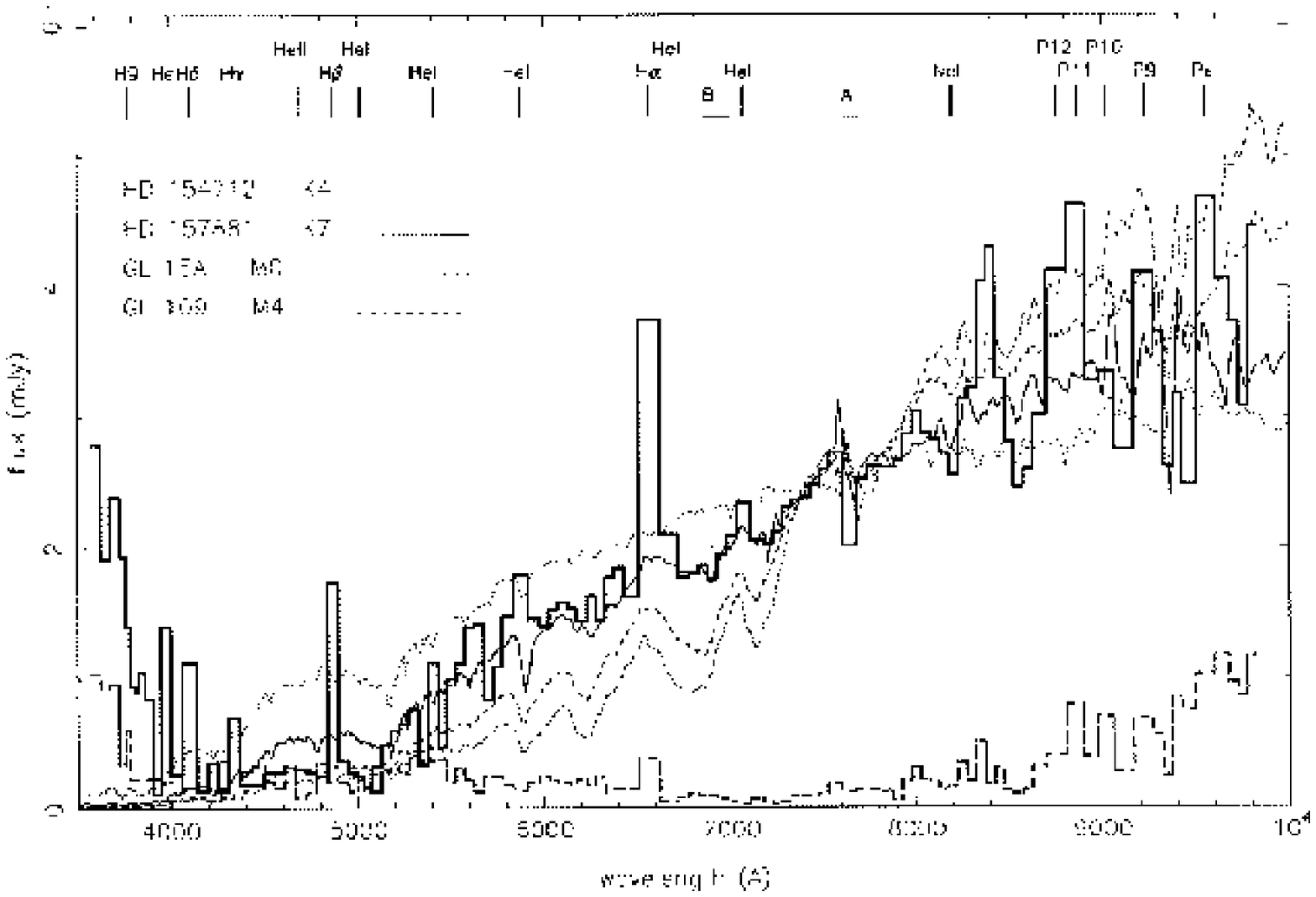} }%
\quad
\scalebox{.5}{%
\includegraphics[bb=1.7cm 2.6cm 19.6cm 24.4cm,angle=-90,scale=.54]
{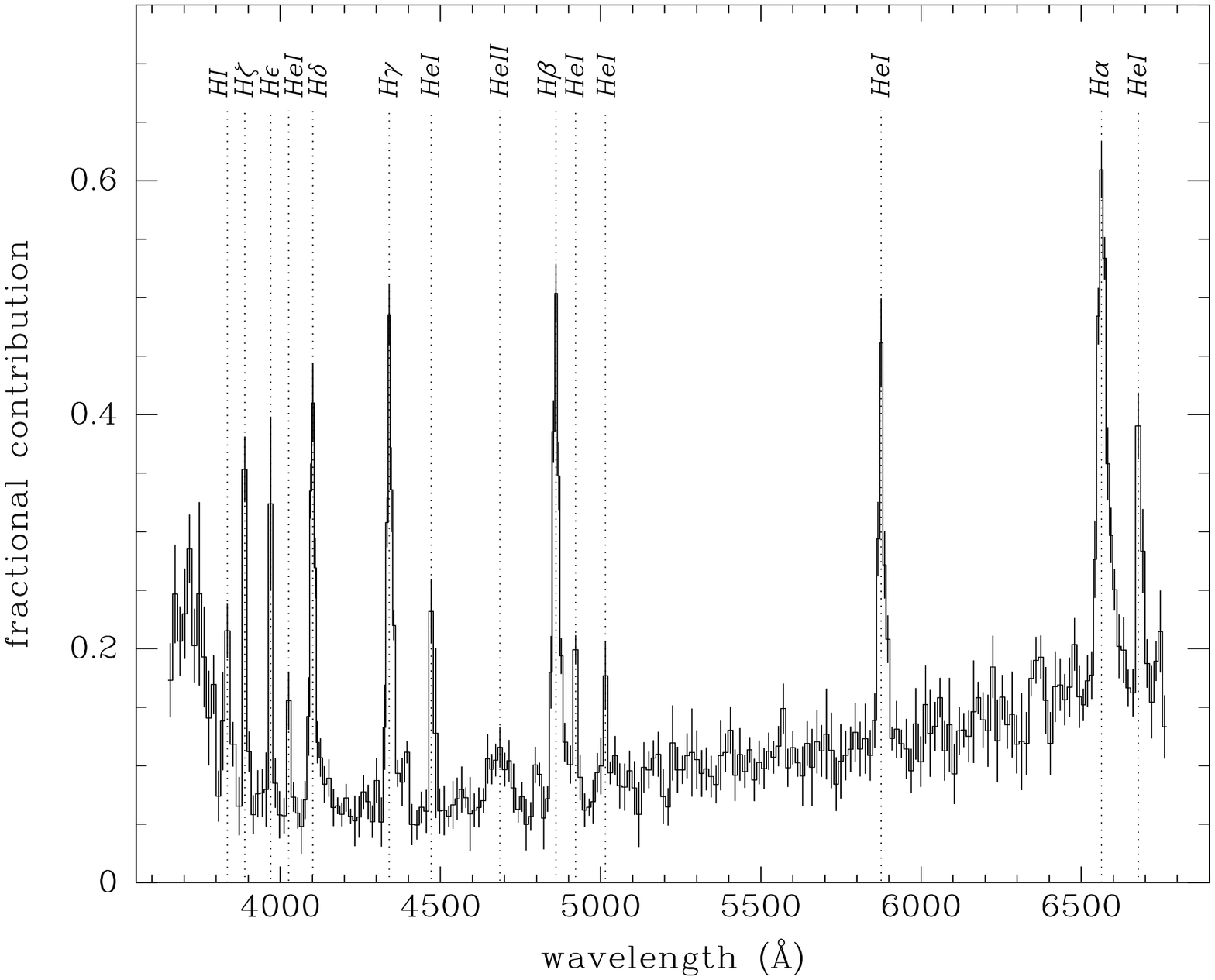} }
\caption{ The spectrum of the uneclipsed component in UX~UMa (left)
  and UU~Aqr (right, expressed as the fractional contribution to the total
  light). From \cite{r94,bssh}. }
\label{fig9}
\end{figure}

The spectrum of the uneclipsed component in these nova-like systems 
shows strong emission lines and the Balmer jump in emission indicating 
that the uneclipsed light has an important contribution from optically 
thin gas (Fig.~\ref{fig9}). 
The lines and optically thin continuum emission are most probably
emitted in a vertically extended disc chromosphere + wind \cite{b98,bssh}.
The uneclipsed spectrum of UX~UMa at long wavelengths is dominated by a
late-type spectrum that matches the expected contribution from the 
secondary star \cite{r94}. Thus, the uneclipsed component seems to provide
an unexpected but interesting way of assessing the spectrum of the 
secondary star in eclipsing CVs.
%
%%%%%%%%%%%%%%%%%%%%%%%%%%%  FIGURE 9  %%%%%%%%%%%%%%%%%%%%%%%%%%%%%%
%%%%%%%%%%%%%%%%%%%%%%%%%%%%%%%%%%%%%%%%%%%%%%%%%%%%%%%%%%%%%%%%%%%%
%

Based on their spatially-resolved results, \cite{b98} 
suggested that the reason for the long standing discrepancies between 
the prediction of the standard disc model and observations of accretion
discs in nova-like variables (e.g., \cite{knigge97,ladous,w84}) is not 
an inadequate treatment of radiative transfer in the disc atmosphere, 
but rather the presence of additional important sources of light 
in the system besides the accretion disc (e.g., optically thin 
continuum emission from a disc wind and possible absorption 
by circumstellar cool gas).

\subsection{Spatial studies} \label{spatial}

Eclipse mapping has also been a valuable tool to reveal that real
discs have more complex structures than in the simple axi-symmetric model.

Besides the normal outbursts, short-period dwarf novae (SU~UMa stars) exhibit 
superoutbursts in which superhumps develop with a period a few per cent
longer than the binary orbital period. Normal superhumps appear early
in the superoutburst and fade away by the end of the plateau 
phase. Late superhumps, displaced in phase by roughly $180\degr$ with 
respect to the normal superhumps, appear during decline and persist into
quiescence  \cite{bible}.  Eclipse mapping experiments have been 
fundamental in testing superhump models.

Reference \cite{o90} analyzed light curves of Z~Cha during superoutburst
with a modified eclipse mapping technique. 
Assuming that the superhump profile is fairly stable over a timescale
of a dozen of binary orbits, he separated the eclipse of the superhump 
source by subtracting the light curve when the superhump maximum occurs
far from eclipse from that in which the superhump is centred on
the eclipse.
The eclipse mapping of the resulting light curve show that the 
superhump light arises from the outer disc, and appears to be 
concentrated in the disc region closest to the secondary star.
This result helped to establish the superhump model of \cite{whit88}, 
in which the normal superhumps are the result of an increased tidal 
heating effect caused by the alignment of the secondary star and an 
slowly precessing eccentric disc.

Further evidence in favour of the existence of eccentric discs in CVs
comes from the recent study of permanent superhumps in the short-period 
novalike variable V348~Pup by \cite{rolfe00}. 
Their eclipse mapping analysis shows that the size of the disc emission
region depends on superhump phase, and that the disc light centre is
on the back side of the disc at superhump maximum.
This phasing is reminiscent of that of the late superhumps in SU~UMa 
stars. Their results indicate that the superhump 
maximum occurs when the secondary star is ligned up with the smallest 
disc radius, suggesting that these superhumps are the result of a
modulation of the bright spot emission caused by the varying kinetic 
energy of the gas stream when it hits the disc edge \cite{vogt81}.

\begin{figure}
\includegraphics[bb=-2.5cm 5.4cm 20cm 20.95cm,clip,scale=0.45]{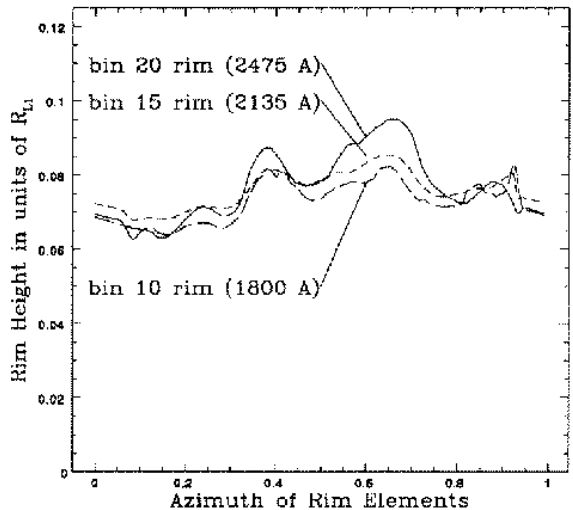}
\caption{The dependency of the disc thickness with azimuth at three 
  different wavelengths for OY~Car in superoutburst. From \cite{bill96}. }
\label{fig10}
\end{figure}

Ultraviolet observations of the dwarf nova OY~Car in superoutburst
show dips in the light curve coincident in phase with the optical
superhump. 
Reference \cite{bill96} analyzed this data set with a modified version
of the eclipse mapping method which simultaneously map the brightness 
distribution on the surface of the accretion disc and the vertical and
azimuthal extent of a flaring at the edge of the disc. 
Their analysis indicates the presence of an opaque disc rim, the thickness
of which depends on the disc azimuth and is large enough for the rim to
obscure the centre of the disc at the dip phase (Fig.~\ref{fig10}). 
These results are consistent with a model of normal superhumps 
as the consequence of time-dependent changes in the thickness of the 
edge of the disc, resulting in obscuration of the ultraviolet flux 
from the central regions and reprocessing of it into the optical part
of the spectrum.
Another evidence of discs with thick rims comes from the work of 
\cite{rob99}, who found that a relatively large disc opening angle is
required in order to explain the ultraviolet eclipse light curves of 
Z~Cha in outburst. It seems that the discs of dwarf novae become flared
during normal outbursts and that the thickening during superoutbursts 
may be sufficient for the disc rim to obscure the inner disc regions.
%
%%%%%%%%%%%%%%%%%%%%%%%%%%%  FIGURE 10  %%%%%%%%%%%%%%%%%%%%%%%%%%%%%%
%%%%%%%%%%%%%%%%%%%%%%%%%%%%%%%%%%%%%%%%%%%%%%%%%%%%%%%%%%%%%%%%%%%%
%

\begin{figure}
\begin{center}
\includegraphics[bb=3cm 9cm 18cm 20.7cm,scale=0.85,clip]{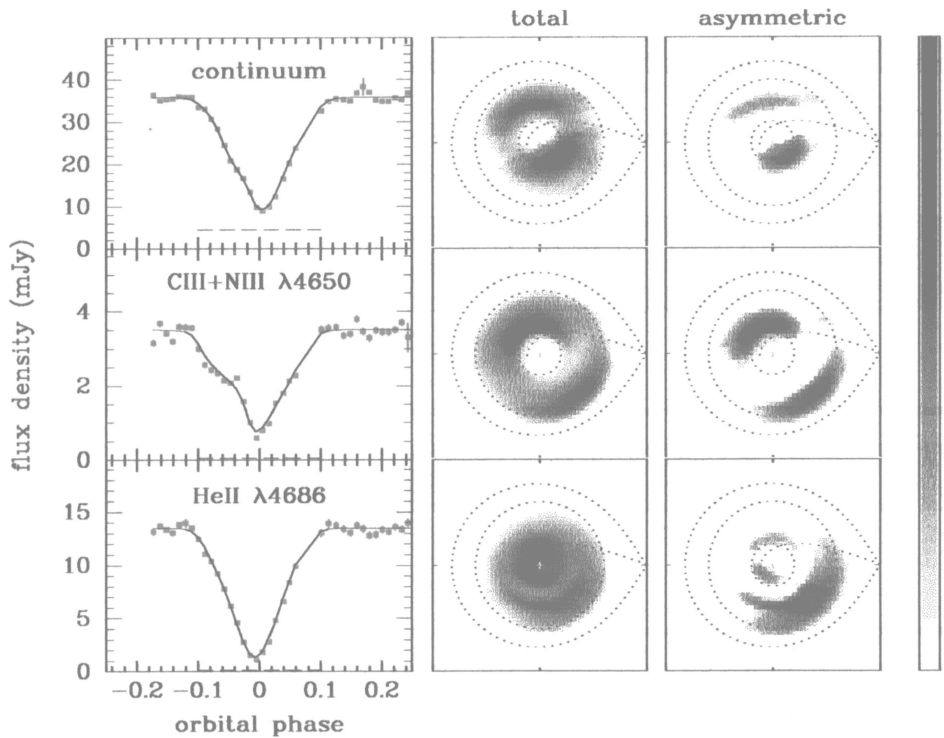}
\end{center}
\caption[]{ Eclipse mapping of spiral shocks in IP~Peg. The light curves
  are shown in the left-hand panels and the eclipse maps are displayed
  in the middle and right-hand panels in a logarithmic greyscale. The
  notation is the same as in Fig.~\ref{simul}.  From \cite{bhs}. }
\label{spiral}
\end{figure}

Tidally induced spiral shocks are expected to appear in dwarf novae 
discs during outburst as the disc expands and its outer parts feel more 
effectively the gravitational attraction of the secondary star 
\cite{makita,steeghs}.
Eclipse mapping of IP Peg during outburst \cite{bhs} helped to 
constrain the location and to investigate the spatial structure of the
spiral shocks found in Doppler tomograms \cite{h99,steeghs97}.
The spiral shocks are seen in the continuum and C\,III+N\,III $\lambda 
4650$ emission line maps as two asymmetric arcs of $\sim 90$ degrees 
in azimuth extending from intermediate to the outer disc regions 
(Fig.~\ref{spiral}).
The He\,II $\lambda 4686$ eclipse map also shows two asymmetric arcs
diluted by a central brightness source.
The central source probably corresponds to the low-velocity
component seen in the Doppler tomogram and is possibly related to 
gas outflow in a wind emanating from the inner parts of the disc.
The comparison between the Doppler and eclipse maps reveal that the 
Keplerian velocities derived from the radial position of the shocks 
are systematically larger than those inferred from the Doppler tomography
indicating that the gas in the spiral shocks has sub-Keplerian velocities.
This experiment illustrates the power of combining the spatial information
obtained from eclipse mapping with the information on the disc dynamics
derived from Doppler tomography.
%
%%%%%%%%%%%%%%%%%%%%%%%%%%%  FIGURE 11  %%%%%%%%%%%%%%%%%%%%%%%%%%%%%%
%%%%%%%%%%%%%%%%%%%%%%%%%%%%%%%%%%%%%%%%%%%%%%%%%%%%%%%%%%%%%%%%%%%%

\subsection{Time-resolved studies}

Eclipse maps give snapshots of the accretion disc at a given time. 
Time-resolved eclipse mapping may be used to track changes in the disc 
structure, p.ex., to assess variations in mass accretion rate or to
follow the evolution of the surface brightness distribution through a 
dwarf nova outburst cycle.

The observed changes in the radial temperature distribution (and
mass accretion rate) of eclipse maps obtained at different 
epochs in the high viscosity, steady-state discs of the novalike 
variables UX~UMa and UU~Aqr are evidence that the mass transfer 
rate in these system is variable \cite{b95,bsh}.

Eclipse maps of Z~Cha during superoutburst show a bright rim in the outer
disc regions which decreases in brightness relative to the inner regions 
as the superoutburst proceeds and the superhumps fade away \cite{wo88}.
This underscores the indications that the superhumps are sited at the
outer disc rim (section~\ref{spatial}).

Reference \cite{r92a} obtained eclipse maps of the dwarf nova OY~Car along 
the rise to a normal outburst. Their maps show that the outburst starts
in the outer disc regions with the development of a bright ring, while
the inner disc regions remain at constant brightness during the rise.
The flat radial temperature profile of quiescence and early rise changes,
within one day, into a steep distribution that matches a steady-state
disc model for \.{M}$= 10^{-9}\;M_\odot$\,yr$^{-1}$ at outburst maximum.
Their results suggest that an uneclipsed component develops during the
rise and contributes up to $\simeq 15$ per cent of the total light at 
outburst maximum. This may indicate the development of a 
vertically-extended (and largely uneclipsed) disc wind, or that the 
disc is flared during outburst (see section~\ref{spatial}).

Time-resolved eclipse mapping covering the decline of an outburst and
of a superoutburst were obtained, respectively, for IP~Peg \cite{bob97} 
and OY Car \cite{b3}. In both cases the radial temperature distribution 
evolves with the inward traveling of a transition front that leaves 
behind a cool disc ($T_b \simeq 5000-6000$~K) while the temperatures 
at the inner disc remain almost constant at a higher value.
The derived speed of this cooling front is $\simeq 0.14$\,km\,s$^{-1}$
for OY~Car and $\simeq 0.8$\,km\,s$^{-1}$ for IP~Peg.

\begin{figure}[t]
\vspace*{8cm}
\includegraphics[bb=17.6cm 2.3cm 18cm 14cm,angle=-90,scale=.53]{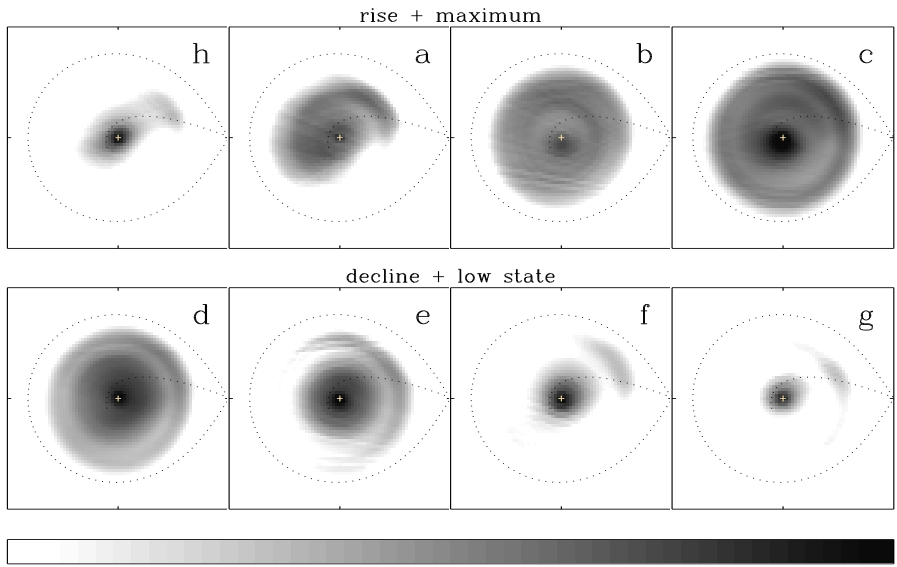}
\caption{ Sequence of eclipse maps of the dwarf nova EX Dra. 
  The eclipse maps capture 'snapshots' of the disc
  brightness distribution in quiescence (h), on the rise to maximum (a-b),
  during maximum light (c), through the decline phase (d-f), and at
  the end of the eruption, when the system goes through a low brightness
  state before recovering its quiescent brightness level. The notation
  is the same as in Fig.~\ref{simul}. From \cite{bc00a}.}
\label{fig12}
\end{figure}

\begin{figure}
\includegraphics[bb=1.7cm 2.5cm 18cm 26cm,scale=.65]{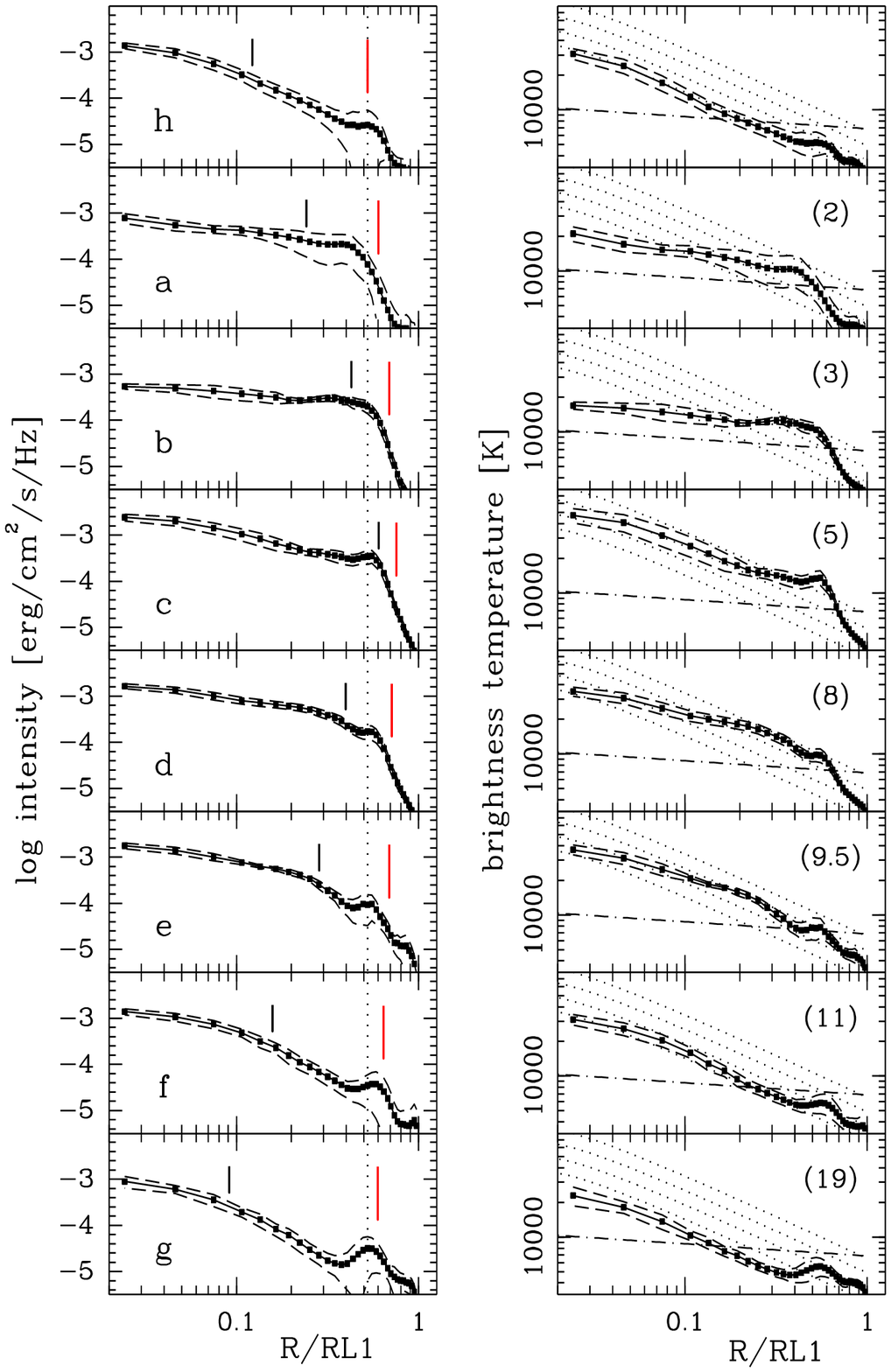}
\caption{ Left: The radial intensity distributions of EX~Dra through the 
 outburst. Labels are the same as in Fig.~\ref{fig12}. Dashed lines
 show the 1-$\sigma$ limit on the average intensity for a given radius.
 A dotted vertical line indicates the radial position of the bright spot
 in quiescence. Large vertical ticks mark the position of the outer edge
 of the disc and short vertical ticks indicate the radial position of a
 reference intensity level.
 Right: The radial brightness temperature distributions. Steady-state disc 
 models for mass accretion rates of $\log$ \.{M}$= -7.5, -8.0, -8.5$, and 
 $-9.0 \;M_\odot\;$yr$^{-1}$ are plotted as dotted lines for comparison. 
 The dot-dashed line marks the critical temperature above which the gas 
 should remain in a steady, high mass accretion regime. The numbers in 
 parenthesis indicate the time (in days) from the onset of the outburst.
 From \cite{bc00a}. }
\label{fig13}
\end{figure}

Eclipse maps covering the full outburst cycle of the long-period dwarf 
nova EX~Dra \cite{bc00a} show the formation of a one-armed spiral 
structure in the disc at the early stages of the outburst \cite{bc00b} 
and reveal how the disc expands during the rise until it fills most of 
the primary Roche lobe at maximum light (Fig.~\ref{fig12}). 
During the decline phase, the disc becomes progressively fainter until 
only a small bright region around the white dwarf is left at minimum light. 
%
%%%%%%%%%%%%%%%%%%%%%%%%%%%  FIGURE 12  %%%%%%%%%%%%%%%%%%%%%%%%%%%%%%
%%%%%%%%%%%%%%%%%%%%%%%%%%%%%%%%%%%%%%%%%%%%%%%%%%%%%%%%%%%%%%%%%%%%
%
The evolution of the radial brightness distribution suggests the 
presence of an inward and an outward-moving heating front during the 
rise and an inward-moving cooling front in the decline (Fig.~\ref{fig13}).
%
%%%%%%%%%%%%%%%%%%%%%%%%%%%  FIGURE 13  %%%%%%%%%%%%%%%%%%%%%%%%%%%%
%%%%%%%%%%%%%%%%%%%%%%%%%%%%%%%%%%%%%%%%%%%%%%%%%%%%%%%%%%%%%%%%%%%%
%
The inferred speed of the outward-moving heating front is of the order of 
1 km\,s$^{-1}$, while the speed of the cooling front is a fraction of that
-- in agreement with the results from IP~Peg and OY~Car.
Their results also suggest a systematic deceleration of both the heating 
and the cooling fronts as they travel across the disc, in agreement with
predictions of the disc instability model \cite{menou}.
A similar effect was seen in OY~Car \cite{b3}.
The radial temperature distributions shows that, as a general 
trend, the mass accretion rate in the outer regions is larger than in 
the inner disc on the rising branch, while the opposite holds during 
the decline branch. Most of the disc appears to be in steady-state at 
outburst maximum and, interestingly, also during quiescence. 
It may be that the mass transfer rate in EX~Dra is sufficiently high to 
keep the inner disc regions in a permanent high viscosity, steady-state.
An uneclipsed source of light was found in all maps, with a steady
component associated to the secondary star and a variable component
that is proportional to the out of eclipse brightness.
Although disc flaring is likely in EX~Dra during outburst, it seems it
is not enough to account for the amplitude of the variation of the 
uneclipsed source.
The variable component was therefore interpreted as emission arising 
from a disc wind, the strength of which depends on the disc mass 
accretion rate.

\section{Future prospects}

Eclipse mapping is a powerful probe of the radial and the vertical disc
structures, as well as of the physical conditions in accretion discs. 
Partly due to the many experiments performed over the last 15 years, 
we have enriched our picture of accretion discs with an impressive 
set of new details such as gas outflow in disc winds, gas stream 
overflow, flared discs with azimuthal structure at their edge,
ellipsoidal precessing discs, sub-Keplerian spiral shocks, and moving 
transition fronts during disc outbursts.

This is however far from being the end of the road. There are still 
many key eclipse mapping experiments remaining to be done.
The spectral mapping of dwarf novae in outburst offers a unique
opportunity to probe the physical conditions of disc spiral shocks
and to critically test the disc instability model by comparing the
spectra of disc regions ahead and behind the transition fronts.
Eclipse mapping estimates of \.{M} for CVs
inside the period gap may be instrumental in testing the current
theories about the origin of the period gap.
A still untouched area is the mapping of the flickering sources.
Our understanding of this fundamental signature of accretion processes 
can certainly be considerably improved with flickering mapping experiments.
The unprecedent combination of high spatial and high spectral resolution
on the disc surface that can be achieved from time-resolved spectroscopic
data yields an unbeatable amount of information to test and improve the
current disc models. Fitting disc atmosphere models to the 
spatially-resolved spectra is the obvious next step to the spectral
mapping experiments and will certainly be rewarding.

Eclipse mapping have been continuously expanding into new domains.
Of promising prospects are
the direct mapping of physical parameters in the accretion disc 
\cite{sonja}, the mapping of the brightness distribution along the 
accretion stream and accretion column in magnetic CVs \cite{kgb00}, 
the 3-D mapping of flared discs and the surface of the secondary star
\cite{dhillon,r98}, and the combination of eclipse mapping and Doppler
tomography \cite{bob99}.

The leap in knowledge about accretion physics that have been (and will
probably continue to be) obtained by the study of close accreting 
binaries with tomographic techniques can lead to a better understanding
of many other astronomical scenarios in which accretion discs may play
an important role, such as AGNs and quasars -- the environment of which
are apparently more complicated and comparatively less well known than 
that of these compact binaries.

\section*{Acknowledgments}

Thanks to Keith Horne and Ren\'e Rutten for stimulating discussions
and valuable advice on the art of eclipse mapping.
CNPq and PRONEX are gratefully acknowledged for financial support 
through grants no. 300\,354/96-7 and FAURGS/FINEP 7697.1003.00.

%INDEX%%%%%%%%%%%%%%%%%%%%%%%%%%%%%%%%%%%%%%%%%%%%%%%%%%%%%%%%%%%%%%%
% Please check with the editor of your book whether he plans to
% include a "mutual" subject index - if so, please code your entries
% in the standard syntax. For your own purposes you may print your
% "personal" index by using the following commands:
%
%\clearpage
%\addcontentsline{toc}{section}{Index}
%\flushbottom
%\printindex
%%%%%%%%%%%%%%%%%%%%%%%%%%%%%%%%%%%%%%%%%%%%%%%%%%%%%%%%%%%%%%%%%%%%%

\end{document}